\documentclass[aps,prl,reprint,superscriptaddress,showpacs]{revtex4-1}
\usepackage{graphicx}
\begin{document}

%\preprint{}

\title{Intrinsic and extrinsic decay of edge magnetoplasmons in graphene}


\author{N. Kumada}
\email[]{kumada.norio@lab.ntt.co.jp}
%\thanks{}
\affiliation{NTT Basic Research Laboratories, NTT Corporation, 3-1 Morinosato-Wakamiya, Japan}
\affiliation{Nanoelectronics Group, Service de Physique de l'Etat Condens\'e, IRAMIS/DSM (CNRS URA 2464), CEA Saclay, F-91191 Gif-sur-Yvette, France}

\author{P. Roulleau}
\author{B. Roche}
\affiliation{Nanoelectronics Group, Service de Physique de l'Etat Condens\'e, IRAMIS/DSM (CNRS URA 2464), CEA Saclay, F-91191 Gif-sur-Yvette, France}

\author{M. Hashisaka}
\affiliation{Department of Physics, Tokyo Institute of Technology, Ookayama, Meguro, Tokyo, Japan}

\author{H. Hibino}
\affiliation{NTT Basic Research Laboratories, NTT Corporation, 3-1 Morinosato-Wakamiya, Japan}

\author{I. Petkovi$\acute{\rm c}$}
\author{D. C. Glattli}
\affiliation{Nanoelectronics Group, Service de Physique de l'Etat Condens\'e, IRAMIS/DSM (CNRS URA 2464), CEA Saclay, F-91191 Gif-sur-Yvette, France}


\date{\today}

\begin{abstract}
We investigate intrinsic and extrinsic decay of edge magnetoplasmons (EMPs) in graphene quantum Hall (QH) systems by high-frequency electronic measurements. 
From EMP resonances in disk shaped graphene, we show that the dispersion relation of EMPs is nonlinear due to interactions, giving rise to intrinsic decay of EMP wavepacket.
We also identify extrinsic dissipation mechanisms due to interaction with localized states in bulk graphene from the decay time of EMP wavepackets.
%To identify extrinsic dissipation mechanisms, we study the time evolution of EMP wavepackets.
%Frequency and temperature dependence of EMP decay time indicates that EMP dissipation is caused through interaction with localized states in bulk graphene.
We indicate that, owing to the unique linear and gapless band structure, EMP dissipation in graphene can be lower than that in GaAs systems.

%The quality factor of our EMP devices is larger than a reported value in GaAs systems.
%We suggest that the lower dissipation is an intrinsic property of graphene, originating from the unique linear and gapless band structure.

%The quality factor of our EMP resonators is larger than a reported value in GaAs systems, stimulating quantum transport experiments and plasmonic applications using graphene.
\end{abstract}

\pacs{a}
%\keywords{}

\maketitle

Edge channels (ECs) in quantum Hall (QH) states provide unique chiral one-dimensional systems to perform a variety of electron quantum experiments including electronic Mach-Zehnder interferometry \cite{Ji,Roulleau} and flying qubits \cite{Neder,Bocquillonscience}.
ECs support collective excitations called edge magnetoplasmons (EMPs) \cite{Volkov,Andrei}, which form the bosonic modes in a Tomonaga-Luttinger liquid representation \cite{Wen}.
EMPs lead to charge fractionalization \cite{Bocquillon,KamataNaturenano} and their dissipation causes energy relaxation \cite{Degiovanni,Levkivskyienergy} and decoherence \cite{Bocquillonscience,Levkivskyi,Wahl} in ECs.
To better understand these physics and to obtain robust quantum effects, investigation of EMP decay mechanism is essential.
EMPs in graphene are of particular interest because, owing to the linear and gapless band structure of graphene, their dissipation would be lower than other conventional two-dimensional systems.
%Furthermore, since the propagation of plasmons and their dissipation are the direct consequence of interactions between charge carriers, plasmon measurements are important for understanding of many-body physics in graphene \cite{YanDamping}.
However, although EMPs in graphene have been observed \cite{Crassee,Yan,KumadagrapheneEMP,Petkovic}, fundamental quantities such as dispersion and decay are yet to be measured.
Furthermore, to the best of our knowledge, the decay mechanism of EMPs in any two-dimensional systems has never been identified.

In this Letter, we report the measurement of intrinsic and extrinsic decay of EMPs in graphene using electronic techniques with the frequency range up to $65$\ GHz.
We prepared disk shaped graphene devices and measured high-frequency transmission in frequency and time domain.
Frequency domain measurement gives resonant frequencies of EMPs, from which the dispersion relation is obtained.
EMP decay time $\tau$ is directly measured in time domain.
From the frequency and temperature dependence of $\tau $, we identify that dissipation is caused by interaction with localized states in bulk graphene.
The quality factor $Q\equiv \pi f\tau $ of our EMP devices is larger than a reported value in GaAs systems \cite{Ashoori}.
We suggest that the lower decay is an intrinsic property of graphene, stimulating quantum transport experiments and plasmonic applications \cite{Viola} using graphene.

\begin{figure}[t]
\begin{center}
\includegraphics[width=0.8\linewidth]{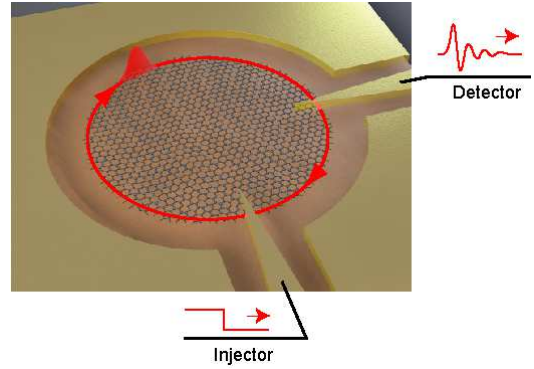}
\caption{
Schematic of the device.
Disk shaped graphene is covered with 160\ nm thick insulating layer.
The capacitive injector and the detector were deposited on top of the insulating layer.
The overlap between the injector (detector) and graphene are 3\ $\mu $m in width.
A sinusoidal wave or a voltage step is sent to the injector and and the current response is detected through the detector.
}
\label{Fig1}
\end{center}
\end{figure}

\begin{figure*}[!t]
\begin{center}
\includegraphics[width=1.0\linewidth]{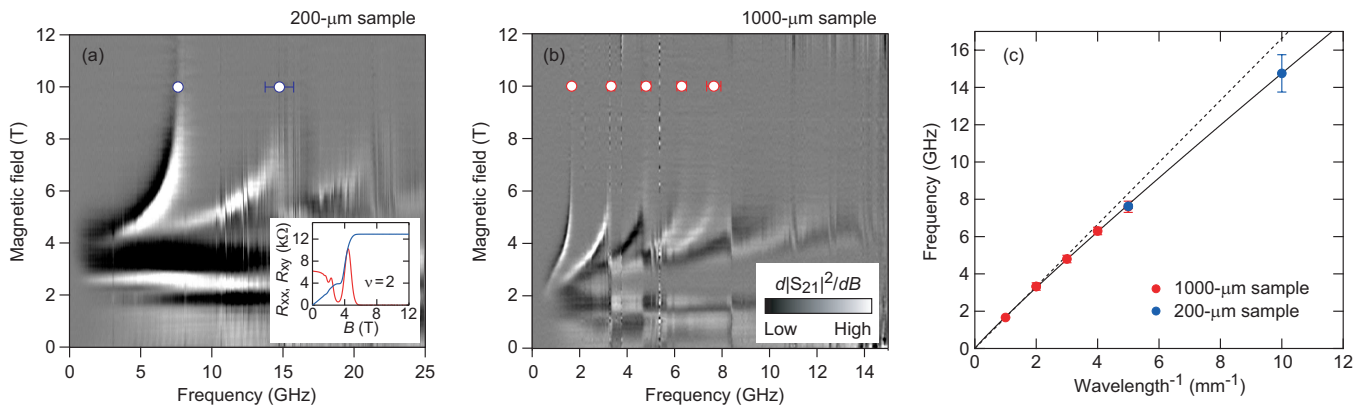}
\caption{Results of frequency domain measurement.
(a) and (b) Transmission signals of the 200- and 1000-$\mu $m samples, respectively.
The differentiated transmission power ($d|{\rm S_{21}}|^2/dB$) is plotted as a function of the frequency and $B$.
%For larger $B$, the $B$ dependence of the signal is weak, resulting in the faint $d|{\rm S_{21}}|^2/dB$.
Open dots with error bar indicate the resonant frequency at high $B$ (details for the accurate determination of the resonant frequency are given in \cite{supplement}).
Inset in (a) shows the longitudinal resistance ($R_{xx}$; red trace) and the Hall resistance ($R_{xy}$; blue trace) of a Hall bar device fabricated from the same graphene wafer obtained by standard DC measurement.
(c) Dispersion relation of the EMP mode: the resonant frequencies [open dots in (a) and (b)] are plotted as a function of $\lambda ^{-1}$ determined by Eq.\ (1).
The data point for the 5th harmonics in the 1000-$\mu $m sample coincides with that for the fundamental mode in the 200-$\mu $m sample at $\lambda ^{-1}=5$\ mm$^{-1}$.
The solid curve is the dispersion relation given by Eq.\ (2).
The dashed line is the result of the linear fit for small $\lambda ^{-1}$ regime, which gives the group velocity $v_g=df /d\lambda ^{-1}=1.7\times 10^{6}$\ m/s.
}
\label{Fig2}
\end{center}
\end{figure*}

\begin{figure}[!t]
\begin{center}
\includegraphics[width=0.8\linewidth]{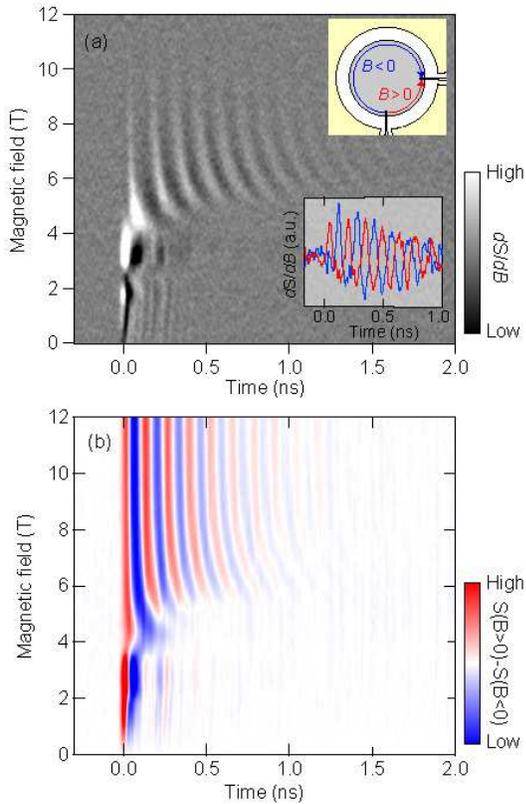}
\caption{
Results of time domain measurement.
(a) Differential signal ($d{\rm S}/dB$) of the 200-$\mu $m sample a function of time and $B$.
Inset shows $d{\rm S}/dB$ at $B=6.5$\ T (red trace) and $B=-6.5$\ T (blue trace).
%The sign of $B$ is defined as positive when $B$ is applied from the top of the graphene.
The chirality of the EMP orbital motion is counterclockwise (clockwise) for positive (negative) $B$.
(b) ${\rm S(B>0)}-{\rm S(B<0)}$ for the 200-$\mu $m sample.
% crosstalk can be eliminated by the subtraction, while the EMP signal is twofold enlarged because the phase of the EMP signal differs by $\pi $ depending on the sign of $B$.
%Because of the difference in the chirality, the phase of the signals for positive and negative $B$ differs by $\pi $.
}
\label{Fig3}
\end{center}
\end{figure}

\begin{figure}[!t]
\begin{center}
\includegraphics[width=0.8\linewidth]{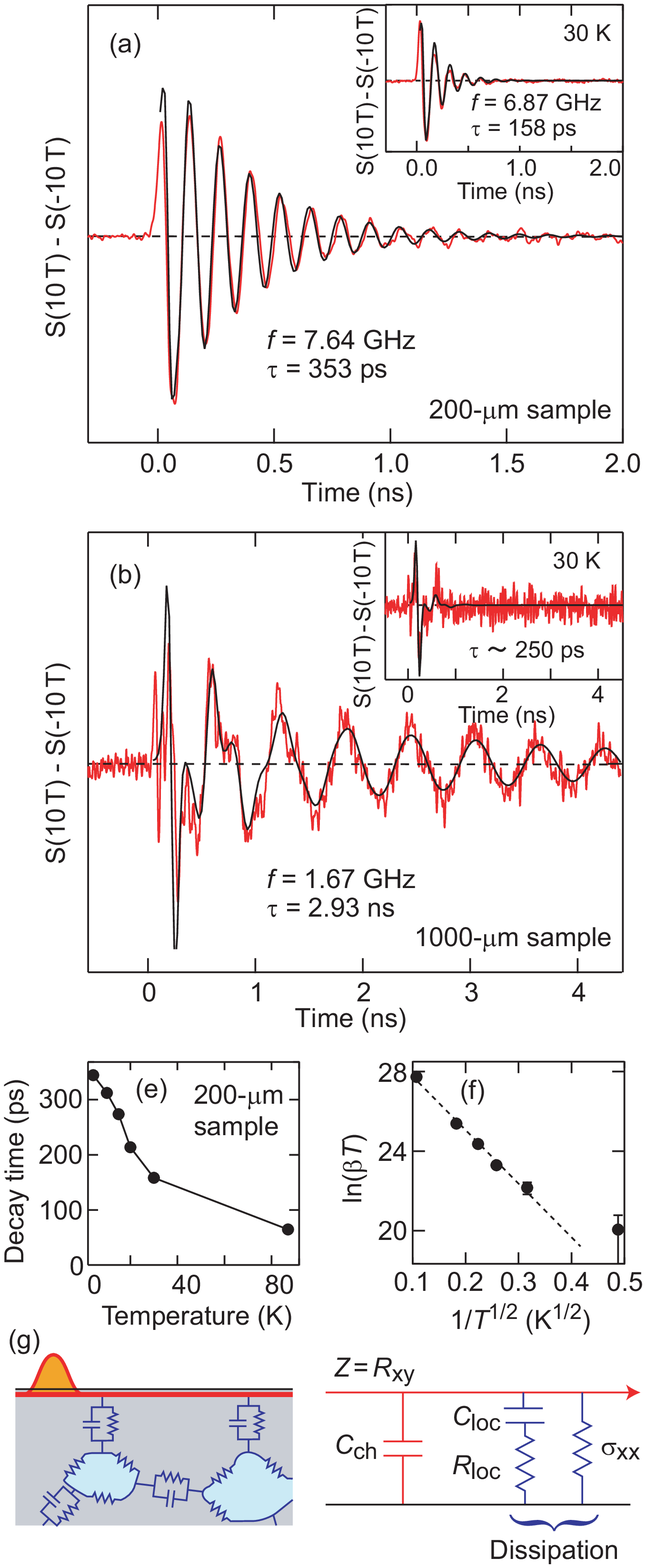}
\caption{
Temperature dependence of EMP decay.
(a), (b) ${\rm S(10 T)}-{\rm S(-10 T)}$ at $T=4$\ K for the 200- and 1000-$\mu $m samples, respectively.
Insets show the traces at $T=30$\ K.
Black traces represent results of the simulation (details are indicated in the main text).
%assuming $\tau ^{-1}=\alpha f^2+\beta (T)$; parameters used are $(\alpha ,\beta )=(4.9\times 10^{-12},0)$, $(1.2\times 10^{-11},0)$, $(4.9\times 10^{-12},3.5\times 10^9)$, and $(1.2\times 10^{-11},3.5\times 10^9)$, for (a), (b), (c), (d), respectively.
%The values of $f$ and $\tau $ indicated are those for the fundamental mode.
(c) $\tau $ for the 200-$\mu $m sample as a function of $T$.
(d) ${\rm ln}(\beta T)$ as a function of $T^{-1/2}$.
Linear fit gives $T_0^{1/2}=26.9\pm 1.4$\ K$^{1/2}$ in Eq.\ (3).
(e) Illustration and equivalent circuit model for the EMP dissipation through interaction with localized states in the interior of graphene.
%The EC can be represented by a capacitively coupled channel-ground line with the impedance $Z=R_{xy}$.
%The dissipation can be modeled as a shunt resistor and capacitor.
}
\label{Fig4}
\end{center}
\end{figure}

We used graphene grown by thermal decomposition of a 6H-SiC(0001) substrate (Fig.\ 1) \cite{Tanabe}.
Graphene is etched into a disk shape.
We used two devices with the perimeter $P$ of 200 and 1000\ $\mu $m called 200- and 1000-$\mu $m samples, respectively.
The carrier is electron and the density is about $5\times 10^{11}$\ cm$^{-2}$.
The mobility is about 12,000\ cm$^2$V$^{-1}$s$^{-1}$.
The surface of graphene was covered with 100-nm-thick hydrogen silsesquioxane (HSQ) and 60-nm-thick SiO$_2$ insulating layers.
%The effective dielectric constant is $\epsilon ^{\ast}_{\rm SiC}=9.6$ for the SiC substrate and $\epsilon ^{\ast}_{\rm HSQ}=2.8$ for the HQS insulating layer.
Two high-frequency lines to inject and detect EMPs were deposited on top of the insulating layer, separating the perimeter in the ratio of 3:1.
%The capacitive injector and detector were deposited on top of the insulating layer.
The overlap between the injector (detector) and graphene are 3\ $\mu $m in width.
In a QH effect regime at high perpendicular field $B$, the active length of the injector and the detector corresponds to the width of ECs, which is estimated to be 10\ nm in a steep edge potential.
Then the impedance of the injector and the detector at 10\ GHz is $\simeq 3$\ M$\Omega $.
This is much larger than the impedance of the EC $R _{xy}\sim 10$\ k$\Omega $, hardly contributing to EMP dissipation.
%EMPs are studied in frequency and time domain by sending, respectively, a sinusoidal wave and a voltage pulse to the injector and detecting the transmitted signal.
Magnetic fields up to 12 Tesla have been used.
The base temperature is $T=4.2$\ K.

We first present the frequency domain results, where the transmission power $|{\rm S_{21}}|^2$ is recorded while sweeping the frequency of the sinusoidal voltage applied to the injector at each value of $B$.
Figures\ 2(a) and (b) plot the differential of the signal with respect to $B$ ($d|{\rm S_{21}}|^2/dB$); by differentiating, we reject $B$-independent crosstalk between the high-frequency lines and highlight the signal associated with EMPs.
For $B\gtrsim 6$\ T, EMP resonances, manifested as a dip and peak structure in $d|{\rm S_{21}}|^2/dB$ appear with almost equal frequency spacing.
The spacing is about 7.6 and 1.6\ GHz for the 200- and 1000-$\mu $m samples respectively, roughly inversely proportional to the perimeter.
As $B$ is decreased, the resonances shift to lower frequency and then are quickly damped.

The EMP resonances can be understood as follows.
The $B$ range for the sharp resonances corresponds to the Landau level $\nu =2$ QH state, in which the longitudinal resistance $R_{xx}$ is vanishing [inset of Fig.\ 2(a)].
In this condition, scattering of EMPs by charges in the bulk graphene is suppressed and EMPs orbit along the graphene edge.
Then the resonance occurs when the wavelength of the EMP mode corresponds to \cite{Balaban},
\begin{equation}
\lambda =\frac{P}{j},
\end{equation}
where $j$ is an integer.
EMPs with $\lambda (j=1)$ is the fundamental mode and $\lambda (j\geq 2)$ is the $j$-th harmonics.
From the resonant frequency $f$, the phase velocity is obtained as $v_p=fP/j$.
The decrease in $f$ on the lower $B$ side of the $\nu =2$ QH state is due to the decrease in $v_p$ induced by the bulk resistance \cite{Volkov,Johnson} (See Supplementary Material \cite{supplement}).

Here, we focus on the resonant modes in the $\nu =2$ QH state at high $B$.
In Fig. 2(c), the frequency of the fundamental and harmonic modes are plotted as a function of $\lambda ^{-1}$ determined by Eq.\ (1): the plot represents the dispersion relation.
The dispersion is almost linear for smaller $\lambda ^{-1}$ with the group velocity $v_g=df /d\lambda ^{-1}=1.7\times 10^6$\ m/s [dashed line in Fig. 2(c)] \cite{Petkovic,KumadagrapheneEMP}.
As $\lambda ^{-1}$ is increased, the dispersion deviates from the linear line and $v_g$ decreases gradually.
This behavior can be reproduced by theory \cite{Volkov} for EMPs in a sharp edge potential,
\begin{equation}
f =\left[ \frac{\sigma _{xy}}{2\pi \epsilon ^\ast \epsilon _0}\left({\rm ln}\frac{2}{2\pi \lambda ^{-1}w}+1\right)+v_D\right] \lambda ^{-1},
\end{equation}
where $\sigma _{xy}$ is the Hall conductance, $\epsilon ^{\ast }$ is the effective dielectric constant, $v_D$ is the drift velocity, and $w$ is the transverse width of EMPs.
The logarithmic term coming from interactions is responsible for the nonlinear dispersion.
The best fit [solid curve in Fig. 2(c)] gives $v_D=(5\pm 1)\times 10^5$\ m/s and $\epsilon ^\ast =14.3\pm 1.4$.
The value of $v_D$ is consistent with previous experiments \cite{Petkovic}.
$w$ is calculated to be 4\ nm \cite{supplement}, which is much smaller than $\sim 2$\ $\mu $m in GaAs/AlGaAs heterostructures, where $w$ is determined by the shape of the slowly varying edge potential \cite{Zhitenev,KumadaEMP}.

Next, we show time domain results, where a EMP wavepacket is generated by a voltage step applied to the injector and then the signal $S$ at the detector is recorded as a function of time \cite{Ashoori}.
The rise time of the voltage step is $\Delta _t=11$\ ps.
Slight increase in $\Delta _r$ at the injector being taken into account, the spatial width of the EMP wavepacket is estimated to be $\Delta _s=\Delta _r\times v_p\sim 40$\ $\mu $m.
Figure\ 3(a) shows the time trace of the differential signal $d{\rm S}/dB$ for the 200-$\mu $m sample.
For $B\gtrsim 6$\ T, the data clearly shows the periodic oscillations with decay.
The phase of the signal reflects the chirality of the EMP motion.
Since the injector and the detector separate the perimeter in the ratio of 3:1, a $\pi $ phase shift occurs when the chirality is changed by changing the direction of $B$ [inset of Fig.\ 3(a)].
This $\pi $ phase shift allows us to eliminate the crosstalk without the differentiation: by subtracting the signal for opposite direction but the same amplitude of $B$, the crosstalk is eliminated while the EMP signal is increased twofold.
Figure\ 3(b) shows the result of the subtraction.
From this plot, the frequency and the decay time of the EMP orbital motion can be directly obtained.

Figure\ 4(a) is the cross section at $|B|=10$\ T, from which $f=7.64$\ GHz and the EMP decay time $\tau =353$\ ps are obtained.
The value of $f$ corresponds to that of the fundamental mode [Fig.\ 2(a)].
Similar measurement for the 1000-$\mu $m sample shows $f=1.67$\ GHz and $\tau =2.93$\ ns [Fig.\ 4(b)].
These results show that $\tau $ grows with decreasing $f$.
The $f$ dependence of $\tau $ changes qualitatively for higher $T$.
At $T=30$\ K, $\tau $ becomes smaller $\sim 200$\ ps for both samples [insets of Figs.\ 4(a) and (b)].
Detailed $T$ dependence of $\tau $ in the 200-$\mu $m sample shows that, although $\tau $ incrases with decreasing $T$, it is limited at around 400\ ps [Fig.\ 4(e)].
These results indicate that the primary mechanism of the EMP dissipation changes with $T$: the $f$ dependent term determines the dissipation at low $T$, while the $f$ independent contribution increases with $T$.

The behavior of $\tau $ can be explained by a model based on the coupling to localized states [Fig.\ 4(g)], which are conductive channels along contours of disorder potential \cite{Huckenstein}.
The localized states couple to EMPs capacitively and resistively.
In a circuit representation, the EC is modeled as a unidirectional transmission line with impedance $Z=R_{xy}$ and channel capacitance $C_{\rm ch}$ \cite{Hashisaka}.
The resistive coupling to the localized states corresponds to a shunt resistor with a conductance $\sigma _{xx}\sim R_{xx}/R_{xy}^2$.
The dissipation by the shunt resistor $\tau ^{-1}\propto \sigma _{xx}/C_{\rm ch}$ is almost independent of $f$, while it increases with $T$.
On the other hand, the capacitive coupling can be represented by a shunt capacitor $C_{\rm loc }$.
The effect of charge excitations within the localized states can be included as a series resistor $R_{\rm loc}$.
The dissipation through the capacitive coupling increases with the admittance of the capacitor $2\pi fC_{\rm loc}$; more precisely, the dissipation is $\tau ^{-1}\propto R_{\rm loc}(C_{\rm loc}f)^2/(C_{\rm ch}+C_{\rm loc})$ \cite{supplement}.
Since $C_{\rm loc}$ is determined by the geometry independent of $T$, this term becomes important at low $T$.
To confirm our model, we simulated the time evolution of the EMP wavepacket \cite{supplement} assuming $\tau ^{-1}=\alpha f^2+\beta (T)$.
In the simulation, the dispersion relation obtained by frequency domain measurements is included.
The simulation with $\beta =0$ reproduces the experimental results at $T=4$\ K [black traces in Figs.\ 4(a) and (b)], indicating that the capacitive coupling is the main source of the decay at low $T$.
The $\alpha $ values used are $4.9\times 10^{-12}$ and $1.2\times 10^{-11}$ for the 200- and 1000-$\mu $m samples, respectively.
The difference of $\alpha $ values would be due to sample dependence of $C_{\rm loc}$ and $R_{\rm loc}$.
Data for higher $T$ can also be reproduced with temperature dependent $\beta $ and the constant $\alpha $ [black traces in the insets of Figs.\ 4(a) and (b)].
The $T$ dependence of $\beta $ follows the variable range hopping law \cite{Efros},
\begin{equation}
\beta \propto \frac{1}{T}\exp [-(T_0/T)^{1/2}],
\end{equation}
with a characteristic temperature $T_0\sim 700$\ K for the hopping between localized states [Fig.\ 4(f)].
This value is consistent with that obtained by the temperature dependence of $\sigma _{xx}$ \cite{Bennaceur}, supporting that the $T$ dependent term comes from the resistive coupling.
It is worth noting that the fine structures that appear in the trace for the 1000-$\mu $m sample are due to nonlinear dispersion.
The width of the injected EMP pulse of $\sim 40$\ $\mu $m is much smaller than the perimeter of 1000\ $\mu $m.
Then the EMP wavepacket is deformed during the propagation until the width reaches the perimeter.

We obtained the quality factor of the resonator $Q=8.5$ and 15.4 at $T=4$\ K for $f =7.64$ and 1.67\ GHz, respectively.
These values are larger than $Q\sim 3.8$ obtained by similar time domain measurements using 2DESs in GaAs/AlGaAs heterostructures with much higher mobility of $6.2\times 10^6$\ cm$^2$/Vs at $f\sim 300$\ MHz and $T=0.3$\ K \cite{Ashoori}.
We suggest that the smaller decay is an intrinsic property of graphene.
Larger cyclotron gap arising from lighter effective mass suppresses the resistive coupling to the localized states.
At the same time, it reduces the size of the localized states, reducing the capacitive coupling.
Atomically sharp edge potential would also contribute to the smaller decay; narrower $w$ at the sharp edge potential prevents EMPs from being excited inside ECs to acoustic charge modes \cite{Aleiner}.
Our results indicate that graphene ECs provides a platform for robust quantum effects, stimulating the use of graphene for quantum transport experiments and plasmonic applications.

\begin{acknowledgments}
The ERC Advanced Grant 228273 MeQuaNo is acknowledged.
The authors are grateful to F. I. B. Williams, S. Tanabe and for experimental support of P. Jacques and M. Ueki.
\end{acknowledgments}

\bibliography{D}

\end{document}